\documentclass[twocolumn, nofootinbib, showpacs, amssymb, amsmath,  prd, superscriptaddress]{revtex4}
\usepackage{epsfig}
\usepackage{graphics}
\usepackage{graphicx}
\usepackage{dcolumn}
\usepackage{bm}
\usepackage{epstopdf}
\usepackage{subfigure}

\newcommand{\be}{\begin{equation}}
\newcommand{\ee}{\end{equation}}

\def\be{\begin{equation}}
\def\ee{\end{equation}}
\def\bea{\begin{eqnarray}}
\def\eea{\end{eqnarray}}

\newcommand{\ve} {{\varepsilon}}

\def\v1{\vspace{1cm}}

\def\ve{\varepsilon}

\usepackage{graphicx}

\usepackage{alltt}

\begin{document}

\title{Baryo-Leptogenesis induced by modified gravities in the primordial Universe}

\author{Liberato Pizza}
\affiliation{Dipartimento di Fisica, Universit\`a di Pisa, Largo B. Pontecorvo, 3, 56127, Pisa, Italy.}
\affiliation{Istituto Nazionale di Fisica Nucleare (INFN), Sez. di Pisa, Largo B. Pontecorvo, 3, 56127, Pisa, Italy.}

\begin{abstract}
The long-standing problem of the asymmetry between matter and antimatter in the Universe is, in this paper, analysed in the context of the modified theories of gravity. In particular we study  two models of $f(R)$ theories of gravitation that, with the opportune choice of  the free parameters, introduce a little perturbation to the scale factor of the Universe in the radiation dominated (RD) phase predicted by general relativity (GR), i.e., $a(t)\sim t^{1/2}$. This little perturbation generates a Ricci scalar different by zero, i.e., $R\neq 0$ that  reproduces the correct magnitude for the asymmetry factor $\eta$ computed in the frame of the theories of the gravitational baryogenesis and gravitational leptogenesis. The opportune choice of the free parameters is discussed in order to obtain results coherent with  experimental data. Furthermore, the form of the potential $V$, for the scalar-tensor theory
conformally equivalent to the $f(R)$ theory which reproduces the right asymmetry factor, is here
obtained.
\end{abstract}


\maketitle

\section{Introduction}
Observational data  suggest that our Universe is composed for the most part of matter, while the antimatter is only presents in trace amounts \cite{Kolb}. Recent studies  propose that the origin of this asymmetry between matter and antimatter lies in the beginning phase of the Universe, before  of the Big Bang Nucleosynthesis (BBN) \cite{Kolb,  Fukugita, Kaplan, brane and baryo, QGB}. Different theories introduce different interactions Beyond the Standard Model (BSM) in order to explain the origin of this asymmetry in the primordial Universe \cite{Kolb,   Fukugita, Kaplan, brane and baryo, QGB}. In this paper we will show how a $f(R)$ theory and a particular non minimal coupling between Ricci scalar and matter are able to explain the cause of this asymmetry. In particular we have shown how a small correction to the standard Hilbert-Einstein action allows the reproduction of a small variation of the scale factor of the Universe in order to reproduce the expected asymmetry factor.  We will describe the $ln(R)$ gravities never introduced before in this context which represent a suitable alternative theory in  order to describe  the early Universe phenomenology.
This work is organised as follows.
In  section II we will briefly introduce some important parameters related to  baryogenesis and to leptogenesis.
In section III we will briefly resume the main topic of  metric $f(R)$ theories of gravity and their implications for  Universe dynamics.
In section IV we will show how modified gravities can reproduce the correct  baryon asymmetry factor by  means of two functional form of $f(R)$ proposed for the first time in this context.
In section V we will  introduce new results about leptogenesis for an already studied form of $f(R)$  and for a new one, i.e. (\ref{frodin}), originally proposed (in this context) for the first time in this paper.
In section VI we will obtain the functional forms of the potential of a  primordial scalar field which, in a scalar-tensor theory of gravity, can realize the lepton asymmetry of the same magnitude of that one generated by the $f(R)$ analysed here. The technique, adopted in order to obtain this potential form, is based on the conformal equivalence between $f(R)$ theories and scalar-tensor ones. Free parameters of this potential are fixed by fixing free parameters of the $f(R)$ theories in order to obtain the expected asymmetry factor. In Section VII we will summarize and comment on our results. 

\section{ Matter-antimatter asymmetry}
In the longstanding list of attempts to explain matter-antimatter asymmetry in the Universe, several parameters have been introduced.
 An important parameter used to quantify the amount of baryon matter that exceeds antibaryon matter is the asymmetry factor $\eta_B$, i.e. the Baryon Asymmetry Factor (BAF):
\begin{equation}
\eta_B= \frac{n_B - n_{\bar{B}}}{s},
\end{equation} 
where $n_B (n_{\bar{B}})$ is the number of baryons (antibaryons)per volume unity  and $s$ the entropy density for the Universe. 
Some works about Cosmic Microwave Background (CMB) anisotropies and BBN show that this factor is $\eta \approx 10^{-10}$ \cite{Kolb}.
Analogously, it is possible to introduce the $\eta$ factor for leptons, i.e. the Lepton Asymmetry Factor (LAF):
\begin{equation}
\eta_{L}= \frac{n_L- n_{\bar{L}}}{s},
\end{equation}
where $n_L (n_{\bar{L}})$  is the number of leptons (antileptons) per volume unity and $s$ the entropy density for the Universe \cite{Kolb}. For $\eta_L$ there are not experimental constraints, but only deductions that estimate it with the same magnitude of $\eta_B$ \cite{Kolb}.
Another useful quantity is the baryons  to photons ratio:
\begin{equation}
n_B/n_{\gamma}\approx 6 *10^{-11},
\end{equation}
or the ratio between quarks and antiquarks in the primordial Universe ($t<10^{-6} s$):
\begin{equation}
\frac{n_q-n_{\bar{q}}}{n_q}\approx 3*10^{-8},
\end{equation}
where $n_q$, $n_{\bar{q}}$ and  $n_{\gamma}$ are respectively the density of quarks, antiquarks and photons in the primordial Universe. 
\section{$f(R)$ theories of gravity}
Since the discovery of the current accelerated phase of the Universe \cite{1998}, and the hypothesis of the early time inflation \cite{Guth}, many alternative models of classical or quantum gravity have been proposed.
$f(R)$ theories are one of the most significant attempts to explain the current expansion of the Universe, the Dark Matter behaviour,  and inflation \cite{Faraoni1, defelix, pizza, review2, reviewona, capozcurv, Odinstov, unphantcosm}. In this paper we will show how $f(R)$ theories can be responsible for the asymmetry between matter and antimatter. $f(R)$ theories are obtained  by replacing the Ricci scalar $R$ in terms of a generic  function $f(R)$, modifying correspondingly the Einstein-Hilbert action $S$ as
\begin{equation} \label{actionmetric}
S=\frac{1}{2\kappa} \int d^4x \,
\sqrt{-g}\, f(R)+S_{M},
\end{equation}
where $\kappa=\frac{8 \pi}{M_P^2}$, $M_P$ is the Planck mass, $g$ the determinant of the metric tensor involved and $S_M$ the  action for  matter terms.   

In order to obtain  dynamical equations, it is possible to vary with respect to the metric $g_{\mu\nu}$ the Eq. (\ref{actionmetric}), obtaining the fourth order field equations \cite{Faraoni1, defelix}
\begin{equation}\label{metf}
f'(R)R_{\mu\nu}-\frac{1}{2}f(R)g_{\mu\nu}-\left[\nabla_\mu\nabla_\nu -g_{\mu\nu}\Box\right] f'(R)= \kappa
\,T_{\mu\nu},
\end{equation}
where
\begin{equation}\label{set}
T_{\mu\nu}=\frac{-2}{\sqrt{-g}}\, \frac{\delta S_M
}{\delta g^{\mu\nu} }  ,
\end{equation}
or equivalently,  splitting the matter counterpart from curvature contribute \cite{pizza, review2}, i.e. $
G_{\alpha \beta} = R_{\alpha \beta} -  \frac{1}{2} R g_{\alpha
\beta} = T^{(curv)}_{\alpha \beta} + T_{\alpha \beta}$
where as shown in \cite{pizza, review2} we define
\begin{eqnarray}
T^{(curv)}_{\alpha \beta} & = & \frac{1}{f'(R)} \left \{ g_{\alpha \beta} \left [ f(R) - R f'(R) \right ] /2 +\right . \nonumber \\
~ & ~ & \nonumber \\
~ & + & \left . f'(R)^{; \mu \nu} \left ( g_{\alpha \mu} g_{\beta
\nu} - g_{\alpha \beta} g_{\mu \nu} \right ) \right \} \label{eq:
curvstress}
\end{eqnarray}
as the curvature energy-momentum tensor, where $;^{\mu \nu}$ denotes the covariant derivative with respect to the indices $\mu$ and $\nu$.  In this work we consider the  Friedmann metric for the Cosmos:
\begin{equation}
ds^2=-dt^2+a^2(t)\left(\frac{1}{1-k r^2}dr^2+r^2 d\Omega^2 \right),
\end{equation}
where $a(t)$ denotes the scale factor of the Universe, $k$  is the curvature of the space , and $d\Omega^2= d\theta^2 +sin^2 \theta d\phi^2 $.
 Thus it is possible to obtain the modified Friedmann equations, i.e.,  \cite{defelix, capozcurv}
\begin{equation}\label{eq: fried1}
H^2 +\frac{k}{a^2} = \frac{1}{3} \left [ \rho_{curv} +
\frac{\rho_m}{f'(R)} \right ],
\end{equation}
and
\begin{equation}\label{eq: fried2}
-2 \dot{H} - 3H^2- \frac{k}{a^2}=  P_{curv}+ \frac{P_m}{f'(R)},
\end{equation}
where the dot, i.e. $\dot{}$, denotes the derivative with respect to the cosmic time, and $P_m$ and $\rho_m$ are respectively the pressure and the density of all fluids which fill the Universe. Besides the curvature density is defined as
\begin{equation}\label{eq: rhocurv}
\rho_{curv} = \frac{1}{f'(R)} \left \{ \frac{1}{2} \left [ f(R)  - R
f'(R) \right ] - 3 H \dot{R} f''(R) \right \},
\end{equation}
and the barotropic Pressure is denoted by
\begin{equation}
P_{curv} = \omega_{curv} \rho_{curv} \label{eq: pcurv}\,,
\end{equation}
where  the effective curvature barotropic factor is given by
\begin{equation}
\omega_{curv} = -1 + \frac{\ddot{R} f''(R) + \dot{R} \left [ \dot{R}
f'''(R) - H f''(R) \right ]} {\left [ f(R) - R f'(R) \right ]/2 - 3
H \dot{R} f''(R)}\,. \label{eq: wcurv}
\end{equation}
In the following  we will use the positive signature $(-,+,+,+)$ and the  Ricci scalar $R$ will be written in function of the Hubble parameter as \cite{pizza, capozcurv}
\begin{equation} \label{eq:constr}
R = 6  \left( \dot{H} + 2 H^2 + \frac{k}{a^2}\right).
\end{equation}
According to PLANCK result \cite{planck}, in the following we will impose the spatial curvature equal to 0, i.e. $k=0$.\\
If we denote the right side of Eq. (\ref{eq: fried1}) and (\ref{eq: fried2}) respectively as $\rho_{eff}= \frac{\rho_m}{f'(R)} +  \rho_{curv}$ and $P_{eff}= \frac{P_m }{f'(R)}+ P_{eff}$, it is possible to introduce the EoS effective parameter as:
\begin{equation}\label{efff}
\omega_{eff}= \frac{P_{eff}}{\rho_{eff}}= -1 - \frac{2 \dot{H}}{3 H^2}.
\end{equation}

\section{Baryogenesis}
In 1967 Sakharov inferred  three necessary conditions to obtain a net baryon asymmetry \cite{Kolb, Sakha}.\\
These three conditions are:
\begin{itemize}
\item existence of reactions violating baryon number;
\item violation of the C and  CP symmetry;
\item the Universe needs to be out of the thermal equilibrium for a finite period of time.
\end{itemize}
Later studies have shown how it is possible to explain the asymmetry relaxing some of these three conditions. 
For example in 1987 Cohen and Kaplan \cite{Kaplan} proposed a model of spontaneous baryogenesis in which the CP violation and out of equilibrium phase were relaxed and  a dynamic violation of CPT symmetry is introduced. Accordingly, the expanding Universe breaks CPT symmetry, that is restored considering a static Universe.
In this model  an interaction between a scalar field $\phi$ (called ilion and it is an axion-like particle) and the baryon number current $J^{\mu}_B$ is introduced, i.e.,
\begin{equation}\label{eq: spontact}
\frac{1}{f}\partial_{\mu} (\phi J^{\mu}_B),
\end{equation}
where $f$ is an energy scale and it is usually bigger than $10^{13} GeV$.
This term dynamically violates CPT in an expanding Universe.
Considering Noether Theorem and considering $\phi$ field uniform in space component we can rewrite (\ref{eq: spontact}) as
\begin{equation}
\begin{split}
\frac{1}{f}\partial_{\mu} (\phi J^{\mu}_B)=\frac{1}{f}\partial_{\mu}(\phi)J^{\mu}_B+ & \frac{1}{f}(\partial_{\mu} J^{\mu}_B)\phi=\frac{1}{f}\partial_{\mu} (\phi J^{\mu}_B)= \\= \frac{1}{f}\dot{\phi}(n_B- n_{\bar{B}}).
\end{split}
\end{equation}

As explained in \cite{Kaplan}, this interaction generates a different population of barions with respect to antibarions before  the field $\phi$ reaches the minimum point of its potential. In this way   the $\eta$ factor is equal to
\begin{equation}
\eta \approx \frac{\dot{\phi}}{g_*f T},
\end{equation}
where T is the temperature at which  the asymmetry and $g_{*}$, i.e. the total number of freedom grades of all the fields present in the early Universe, are computed. Notice that $T$ is also the temperature to which the reactions, which violate the baryon number, decouple from the background.

\subsection{Gravitational baryogenesis for $f(R)=R+ \alpha R^2$}
Gravitational baryogenesis was first proposed by J.Paul Steinhardt et al.in 2004 \cite{Stein}. Inspired from spontaneous baryogenesis \cite{Kaplan} they proposed this kind of interaction:
\begin{equation} \label{stein}
\frac{1}{M_*^2}\partial_{\mu} (R J^{\mu}_B),
\end{equation}
where $R$ is the Ricci scalar,  $J^{\mu}_B$ is the baryon current,  $M_{*}$ the cut-off scale of the effective theory.
This   interaction emerges in the phenomenology of some theories  of quantum gravity or supergravity \cite{Stein, marstring, witten, confvsbd, Kaku, polchinsky}.
Relaxing  the equilibrium condition it is possible to reproduce the correct asymmetry because interaction (\ref{stein})  dynamically violates CPT. In this case $\eta$ becomes:
\begin{equation}\label{etaa}
\eta=\frac{\dot{R}}{M_*^2 T_D},
\end{equation}
where $T_D$ is the temperature at which  the decoupling of  reactions that violate baryon number from primordial plasma occur. 
Tracing general relativity equations \cite{d'inverno}, i.e. $R_{\mu\nu}-\frac{1}{2}g_{\mu\nu}R=\frac{8 \pi}{M_p^2} T_{\mu \nu}$, we can express Ricci scalar as:
\begin{equation}
R= 8 \pi\frac{(1- 3 w)\rho}{M_p^2}.
\end{equation}
The baryogenesis happened in a radiation dominated phase, when the EoS parameter is $1/3$, i.e. $w=1/3$, so the Ricci and its derivative are equal to zero. If we hypothesize perturbative effects in QFT \cite{Stein}, or a modified gravity we can obtain an adiabatic index slightly different from $1/3$. \\
In the following, we will show how an $f(R)$ theory can reproduce the correct value of the Ricci scalar and its first derivative, in order to explain the expected asymmetry factor.
Indeed, we will search for an $f(R)$ theory of gravity  that can reproduce a scale factor slightly different from the expected one for the radiation dominated phase, i.e.,
\begin{equation}\label{am}
 \tilde{a}(t)= a_0 t^{1/2} +\lambda(t),
\end{equation} 
with $\lambda(t)$ little as much as to not change the known thermal history of the Universe.
 
In the following,  inspired by the work \cite{Lamb, Lamb2}, which first suggested the idea of applying $f(R)$ in order to explain baryogenesis,  we will study a $f(R)$ form which can, differently from the one proposed in \cite{Lamb}, evade the solar system test \cite{soltest, soltest1}, and that is  more suitable to describe the early Universe dynamics, i.e.,
\begin{equation} \label{frmia}
f(R)= R+\alpha R^2,
\end{equation}
where $\alpha$ is a constant having the dimension of $GeV^{-2}$.
In particular, immediately hereafter we will show how this $f(R)$ can generate a little shift of the scale factor, from the standard one, i.e. $a=a_0 t^{1/2}$ in Radiation Dominated (RD) phase, where $a_0$ is a dimensional constant.\\
We are looking for a solution of the scale factor given by Eq. (\ref{am}), and in order to find this form of solution, we substitute the (\ref{am}) expression in the  Friedmann Equations (\ref{eq: fried1}, \ref{eq: fried2}). We solve Friedmann Equations   in the  linear approximation for $\lambda(t)$. 
 This linearization is performed because we suppose that $\lambda(t)$ is a little perturbation with  respect to the scale factor $a(t)$. Indeed we will check the time lapse in which this linearization is true by the means of the introduction of the $\epsilon$ parameter, i.e.
\begin{equation} \label{eps}
\epsilon= \frac{\lambda}{a}= \epsilon_0 (\frac{T}{M_P})^{\gamma},
\end{equation}
with $\epsilon_0= (\frac{4}{3}\pi \sqrt{\frac{\pi g_*}{5}})^{\gamma/2} \frac{\lambda_0 M_p^{\gamma/2}}{a_0}$. In the previous equation we have used  the relationship between cosmic time and temperature in the radiation dominated phase, i.e., $t=(\frac{90}{32 \pi^3 g_*})^{1/2} \frac{M_p}{T^2}$ (with $g_* \approx 106$).
The linear approximation is valid if $\epsilon \ll 1$.
 Besides we solve  Friedmann Equations  imposing $w=1/3 +\delta$ and  the following ansatz for $\lambda(t)$:
\begin{equation}
\lambda(t)= \lambda_0 t^{\beta},
\end{equation}
with $\lambda_0$ dimensional constant.\\

\begin{figure}
\includegraphics [scale=1] {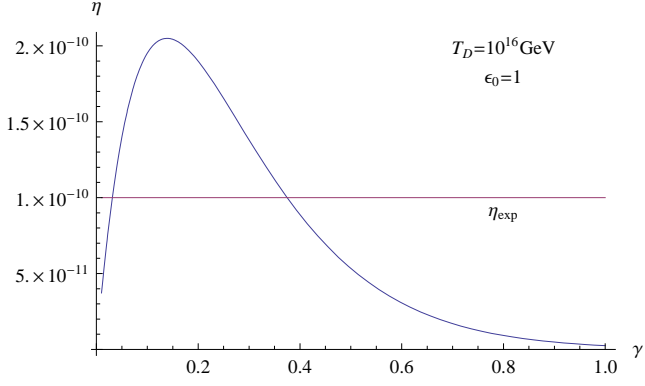}
\caption{\textit{ Baryon  Asymmetry Factor (BAF) in function of $\gamma$.}}\label{eps1}
\end{figure}

\begin{figure}
\includegraphics[scale=1]{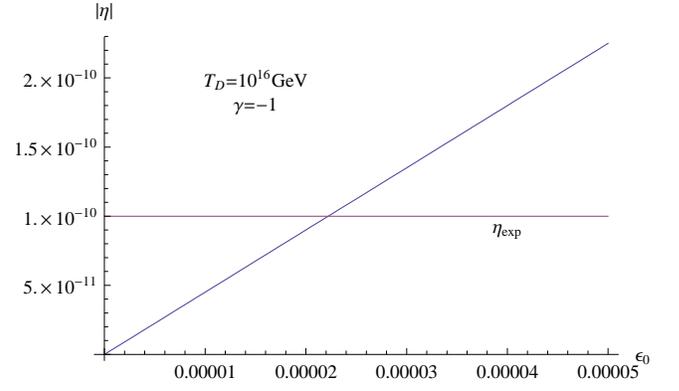}
\caption{\textit{BAF in fucntion of $\ve_0$ for $f(R)=R+\alpha R^2$.   }} \label{eps2}
\end{figure}

In this model the Ricci scalar computed from the modified scale factor $\tilde{a}$, always in the linear approximation, is equal to 
\begin{equation}
R= \frac{3 \lambda}{2 a t^2}(4 \beta^2-1),
\end{equation}
and its time derivative is:
\begin{equation}
\dot{R}= \frac{3}{2}(4 \beta^2 -1) (\beta- \frac{5}{2}) \frac{\lambda}{a t^3}.
\end{equation}
In this way we can substitute this expression in the Eq. (\ref{stein}) and using the relationship between time and temperature in the radiation-dominated phase,  the parameter $\epsilon$ defined in (\ref{eps}) and imposing $M_* \approx M_P$ we  obtain the asymmetry factor as:
\begin{equation}
\eta \approx \frac{3}{4}\epsilon_0 \left(\frac{16 \pi^3g_*}{45}\right)^{3/2}\gamma(2-\gamma)(\gamma+4)\left(\frac{T_D}{M_P}\right)^{\gamma+5}.
\end{equation}
where   $\gamma=1-2\beta$ and $T_D$ is the temperature at which the reactions which violate baryon number decouple from the background. Please note that $\epsilon_0$ is a dimensionless constant because of the different dimensionality of the constant $a_0$ and $\lambda_0$. Given the arbitrariness of the ratio $\lambda_0/ a_0$ we can choose it as small as necessary in order to obtain $\epsilon_0$ equal to $1$ as it will be done in the following.
In Figure \ref{eps1} and \ref{eps2} we report respectively $\eta$ in function of $\gamma$ and $\epsilon_0$,  for a determinate  value of decoupling temperature, i.e. $T_D=10^{16}GeV$. The model (\ref{frmia}) is able to reproduce a correct asymmetry factor for a wide range of $\gamma, \epsilon_0$.
Furthermore the model consider the necessary condition  $\epsilon<<1$. If we suppose 
 $\ve_0$ equal to 1 and $T_D\simeq10^{16}GeV$  we get the results presented below:
\begin{equation}
\epsilon=10^{-3\gamma}\sim\begin{cases} 0.5\:\ \text{per}& \gamma=0.1 \\
0.06\:\ \text{per}&\gamma=0.4 \\
\end{cases}.
\end{equation}



\begin{figure}
\centering
\includegraphics[scale=1]{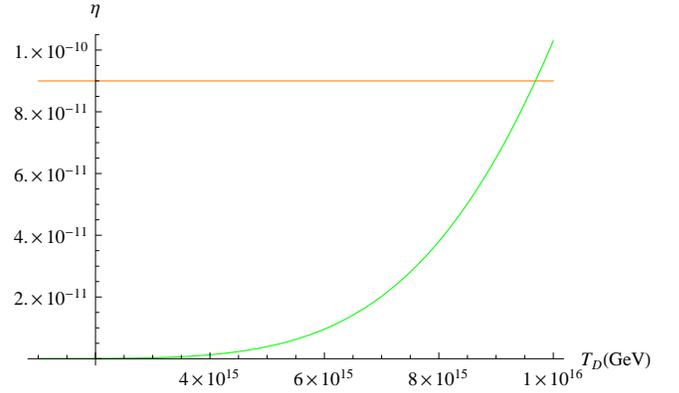}
\caption{\textit{BAF $\eta$ in function of the Temperature
 for   $\delta=0.001 $ (green line) for the model $f(R)=R+\alpha R^2$. Expected value for $\eta$ is in yellow.}}\label{grafic1}
\end{figure}

In Figure \ref{grafic1} we show $\eta$ in function of $T_D$.    
Our model is consistent with observational cosmic history  because it verifies:

\begin{enumerate}
\item $\eta\simeq10^{-10}$,
\item $\epsilon \ll 1$,
\item $\delta \ll 1/3$.
\label{3req}
\end{enumerate}
 
\newpage
\subsection{Gravitational baryogenesis in   $\ln(R)$ theories }
In this section we describe a  $f(R)$ model,  originally proposed in this paper as a suitable attempt to reproduce the expected BAF.  In particular this modified gravity can produce, in a very simple way, a slight variation of the scale factor with respect to the scalar factor of the radiation dominated phase predicted  by General Relativity (GR). We consider again the presence of the interaction (\ref{stein}) that allows the splitting of the energetic level between matter and antimatter. 
The $f(R)$ model in exam is \cite{Odinstov}:

\begin{equation}\label{frodin}
f(R)=R+\gamma R^{-n}(ln \frac{R}{\mu^2})^m,
\end{equation}
where $n$ is a real number, i.e. $n>-1$, $m$ is arbitrary, $\gamma$ and $\mu$ are some dimensional constants.
It is easy to show that the solution of Friedmann Equation for this $f(R)$ is \cite{Odinstov}:
\begin{equation} \label{aodin}
a\backsim t^{\frac{(n+1)(2n+1)}{n+2}},
\end{equation}
with the effective EoS parameter (defined in (\ref{efff})) equal to:
\begin{equation}
w_{eff}=-\frac{6 n^2+7n-1}{3(n+1)(2n+1)}.
\end{equation}
 It is easy to show that for $n=0$ scale the factor has the same behaviour of GR scale factor in the RD phase. Straightforwardly the power-law of the solution depends only on $n$ and not on $m$ and other two dimensional constants. 
 
 Substituting   (\ref{aodin}) in  (\ref{eq:constr}), and deriving the result with respect to the cosmic time, we can get $\dot R$, i.e., 
 \begin{equation}\label{Rpodin}
\dot R=\frac{12 n (5 + 19 n + 22 n^2 + 8 n^3)}{(2 + n)^2 t^3},
\end{equation}
which is equal to $0$ for $n=0$, as expected. Rewriting 
(\ref{Rpodin}) in terms of the temperature,  we can rewrite all as:
\begin{equation} \dot R=\frac{12 n (5 + 19 n +
22 n^2 + 8 n^3)T^6}{(2 + n)^20.027g_*^{-3/2}M_{P}^3}.
\end{equation}
In this way the asymmetry factor  (\ref{etaa}) becomes:
 \begin{equation}
\eta=\frac{12 n (5 + 19 n + 22 n^2 + 8 n^3)T^5}{(2 + n)^20.027g_*^{-3/2}M_{P}^3M_{*}^2} \Bigg|_{T_{D}}.
\end{equation}

\begin{figure}
\centering
\includegraphics[scale=1]{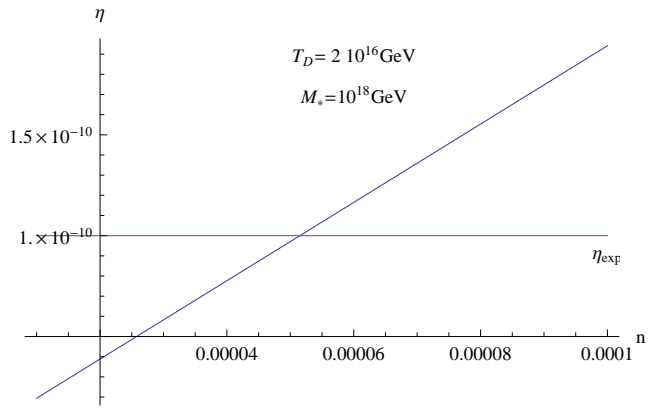}
\caption{\textit{BAF in function of $n$. For $n$ of the order of $10^{-5}$ ($a \sim t^{0.50001}$) we reproduce expected BAF.}} \label{eps4}
\end{figure}

For example, if we choose  $n=0.01$ that reproduces a scale factor proportional to  $t^{0.51}$ and impose decoupling temperature equal to $10^{16} GeV$, we get the following $\eta$ factor
\begin{equation}\label{etaaet}
\eta= \frac{6\cdot10^{26}}{M_*^2}GeV^2.
\end{equation}
We obtain results  with the expected BAF,  adopting  $n$ smaller than $0.01$, which reproduces a scale factor like $t^{0.5001}$ or $t^{0.50001}$, in perfect agreement with the thermal history of the Universe.
 In Figure \ref{eps4} we  show the behaviour of $\eta$  in function of $n $, for some fixed values of  other parameters.

\section{Gravitational leptogenesis}
If we suppose that the Universe has a baryon asymmetry, it is straightforward to speculate about a lepton asymmetry. Common sense leads us to hypothesize that this asymmetry is of the same order of magnitude as the baryon one. In particular the charge neutrality of the Universe is a good evidence to support this assumption, even if one of the main problems is that we do not have any direct measurement about the lepton number given by  three neutrinos. Predictions of BBN assume that neutrino lepton number is very small \cite{Kolb}.

In the last few decades different  models of
leptogenesis  have been developed  \cite{Kolb, Fukugita, RHdecay, Barioleptooo, LambM2, lepto1, leptorew, leptotempesta, mio}. In the following we will consider a theory that explains lepton asymmetry as a consequence of scalar curvature, similarly as we have  proceeded for the gravitational baryogenesis \cite{lepto1, LambM2, mio}.
 In the  model \cite{lepto1, LambM2, mio} the origin of  lepton asymmetry is
realized by a new lagrangian in which  a coupling between Majorana Neutrinos and Ricci scalar. This term violates CP is present.  
Hereafter we will explain the basic idea behind this interaction.
We start  realizing that  in the Standard Model  only neutrinos with defined chirality exist, i.e., 
 left-handed neutrino (LHN)
and right-handed antineutrino (RHA) . Some extensions of the Standard Model introduce
heavy right-handed   neutrino  (HRN) and  heavy  left-handed  antineutrino (HLA). These neutrinos, whose masses are described by the see-saw mechanism \cite{seesaw, neutrino},  are  main actors in the model of leptogenesis studied hereafter. The above introduced interaction, which violates CP, and that it is even under C, and odd under P, is described by \cite{LambM2, mio}: 
 \begin{equation} \label{CPv}
\mathcal{L}_{\diagup{\!\!\!\!\!\!C\!\!P}}=\sqrt{-g} \theta R \bar
\psi i\gamma_{5}\psi,
\end{equation}
where $\theta$ is a constant that introduces the effective range of validity of the model, i.e. $M_P^{-1}$,
$\psi$ is a fermionic field (neutrino in our case), $i$ is the imaginary unit, 
$\gamma_5$ the chirality operator.\footnote{The $\gamma_5$ is
the product of four Dirac matrices:
$\gamma_5=i\gamma_0\gamma_1\gamma_2\gamma_3$. Trough Wick rotation we can recast it in the Euclidean space as
$\gamma_5= \gamma_0\gamma_1\gamma_2\gamma_3$.   Dirac Matrices
allow to build  fermionic fields observables and they
satisfy anticommutation relationship $\{\gamma_\mu,
\gamma_\nu\}=2 Ig_{\mu\nu}$.}\\
The operator   (\ref{CPv}) conserves  CPT
only in a  static Universe and dynamically violates it
in an expanding Universe, i.e.
$\dot R\neq0$.\\
Introducing (\ref{CPv}) the energy of a fermion is determined by its chirality through the interaction with the gravitational background. Euler-Lagrange equations give this result for  dynamics of  fermions coupled non minimally to the background
\footnote{The equation is obtained by varying  the action composed by the
Dirac term plus the term which violates  CP, i.e:
   $\mathcal{ L}= \bar
\psi(i\partial^{\mu}\gamma_\mu -m)\psi+\theta R \bar \psi
i\gamma_{5}\psi$.}

 \begin{equation}
i\gamma^{\mu}\partial_{\mu}\psi-M\psi-i\theta R\gamma_5\psi=0.
\end{equation}

Energetic dispersion relation is given by (see \cite{LambM2, mio} for clarifications):
\begin{equation}E^2\psi=(p^2+M^2+\theta^2R^2)\psi-\theta(\gamma_5\gamma^{\mu}\partial_{\mu}R)\psi.\end{equation}

Now we express $\psi$  as a  superposition of  a left-handed spinor $\psi_{-}$ and of a right-handed one $\psi_{+}$, i.e.
\begin{equation} \psi=\psi_{-}+ \psi_{+},
\end{equation}
where
\begin{equation}
\psi_{+}=\frac{(1+\gamma^5)}{2}\psi , \quad\psi_{-}=\frac{1-\gamma^5}{2}\psi.
\end{equation}
We use this property for $\gamma_5$: \begin{equation}
\gamma^5 \psi_{}= \begin{pmatrix}1 & 0 \\
0 & -1 \\
\end{pmatrix}\begin{pmatrix}\psi_{+} \\
\psi_{-} \\
\end{pmatrix}_{}=\begin{pmatrix} \psi_{+}\\
-\psi_{-} \\
\end{pmatrix}.
\label{chiralmatrix}
\end{equation} 
 It is easy to show that, 
through chiral properties (\ref{chiralmatrix}),  the effective energetic levels of  spinors are split in function of their chirality \cite{LambM2, mio}, i.e.,
\begin{equation}
E^2\psi_{\pm}=(p^2+M^2+\theta^2R^2\pm\theta \dot R)\psi_{\pm.}\label{diracdispersion}
\end{equation}
 From (\ref{diracdispersion})  we get this expression:
 \begin{equation}
 E_{\pm}=\sqrt{{\bf p}^2 +M^2 + \theta^2 R^2}\,  \mp \, \frac{\theta \dot R}{ 2 \sqrt{{\bf p}^2
 +M^2 + \theta^2 R^2}}\,.
 \end{equation}
 If  we consider  Majorana neutrinos in the chiral form  $(N_R, N_R^c)^T$ we can denote  the first component of the bispinor $N_R$ as the HRN while the second one as the HLA, i.e. $N_R^c= \bar N_L$. 
  
If we denote  $N_R$ with $\psi_+$ and  $\bar N_L$ with $\psi_{-} $ the energetic levels of this particle are given by (\ref{diracdispersion}). In particular we point out that effective masses of two neutrinos are different by means of the gravitational interaction. 

Hence from (\ref{diracdispersion}) and for $\theta\dot R \ll M$  we get:

\begin{equation}\label{massaeff}
M_{\pm}=M+\frac{\theta^2R^2}{2M}\pm\frac{\theta\dot R}{2M},
\end{equation}

 that shows how left-handed fermion component has an effective minor mass
respect to the right-handed one. 
Furthermore the rate of decay for a massive heavy neutrino is \cite{LambM2, mio}:
 \begin{equation}
\Gamma_{\pm}=\frac{1}{8 \pi}h^2M_{\pm},
\end{equation} where $\pm$ denotes particle chirality and $h$ is Youkawa coupling.

Lepton asymmetry factor can be witten as \cite{LambM2, mio}: \begin{equation}\label{lamb2} \eta=\frac{\Gamma_+
-\Gamma_-}{\Gamma_+ +\Gamma_-}=\frac{M_+-M_-}{M}=\theta\frac{\dot
R}{M^2},
\end{equation}
in the limit of $\theta R<< M$.  In the following we will consider $\theta \approx M_P^{-1}$.\\Please note that in (\ref{lamb2})  the mass of the heavy neutrino is present. This mass, nowadays, has not a certain value. Different theoretical models proposed a wide range for this mass that goes from $10^{^{7}} \:\ GeV$ to $10^{16}\: GeV$.
Hereafter we will estimate the mass of the heavy neutrino in function of the decoupling temperature, i.e., $T_D$, for the lepton number violation reaction.
 
\subsection{  $ f(R)=R+\alpha R^2$ applied to leptogenesis  }
\begin{figure}
 \includegraphics[scale=1]{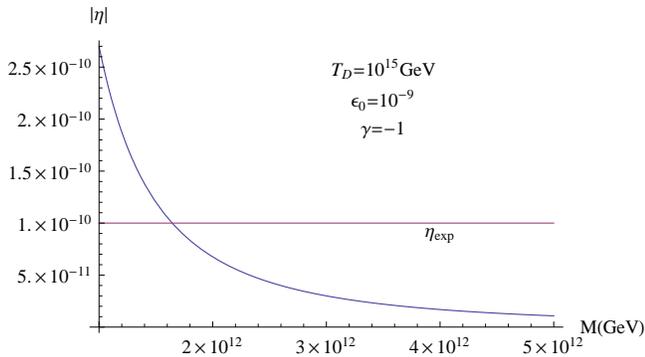}
 \caption{\textit{LAF $\eta$ in function of HRN for $f(R)=R+ \alpha R^2$.}}
 \label{eps3}
 \end{figure}

 In the following, we will compute the correct lepton asymmetry factor (LAF) in  $f(R)=R+ \alpha R^2$ gravity and we will give a probant prediction of the mass of the right-handed neutrino.   If we substitute (\ref{eq:constr}) in (\ref{lamb2}) and we express everything in function of temperature we get the following expression for  LAF: \begin{equation}\label{etalamb2}
\eta= \frac{3}{4}\ve_0 \left(\frac{16 \pi^3g_*}{45}\right)^{3/2}\gamma(2-\gamma)(\gamma+4)\left(\frac{T_D}{M_P}\right)^{\gamma+4}\left( \frac {T_D}{M} \right)^2.
\end{equation}
In Figure \ref{eps3} we reproduce $\eta$ in function of the mass
 M  for $\ve_0=10^{-9}, \gamma=-1 $,  $T_D=10^{15}
GeV$. The correct LAF
$\eta_{exp}\simeq 10^{-10} $ is obtained for $M\sim 10^{12} GeV$.
\\  If we increment the Decoupling temperature (till to $10^{16} GeV)$ we see that the mass of the heavy neutrino necessary to produce the correct LAF without violate BBN constraints \cite{mio}  is of the order of $10^{16} GeV$. 

Now through  see-saw mechanism \cite{seesaw, neutrino} we estimate light neutrino mass from heavy neutrino mass. The relationship between light and heavy neutrino from the see-saw is:
\begin{equation}\label{seesaw}
m_\nu M\simeq m_D^2,
\end{equation}
where $m_{D}$ is the Dirac Mass, i.e.,  the mass of one of  standard model fermions. Considering our best fit for neutrino equal to $M=10^{12} GeV$ and  choosing for $m_D$ quark bottom mass ($4GeV$), we get our prediction for light neutrino mass: $m_\nu \simeq 10^{-1 }eV$ a value in perfect agreement with experimental constraints from neutrino oscillation in the atmosphere \cite{neutrino}. Analogously, if we apply the same mechanism to our second fit for heavy neutrino ($M=10^{15} GeV$), and choosing as $m_D$ quark top mass ( $10^2  GeV$), we get another consistent value for our prediction of light neutrino mass ($m_\nu\simeq10^{-2} eV$). 

At the end we should emphasize a process in which it is possible to obtain baryogenesis via leptogenesis, or rather a baryon asymmetry from a lepton original one. 
In our model we have shown how the gravitational interaction allows a production of an asymmetry between HRN and HLA. This asymmetry is transferred to ordinary light neutrinos via decay processes of the HRN and HLA (for further details see \cite{NdecayW, lepto1}). This happened at $T\sim M \sim 10^{12} GeV$. Note that the necessary condition for our model is $T_D> M$.  Thus, neutrino asymmetry is transferred to  lepton sector through these decays. 
The lepton asymmetry can be converted in the baryon asymmetry during sphaleron era, as shown in \cite{Kolb, sphaler}.  

\subsection{Leptogenesis in $\ln(R)$} 
\begin{figure}
\centering
\includegraphics[scale=1.]{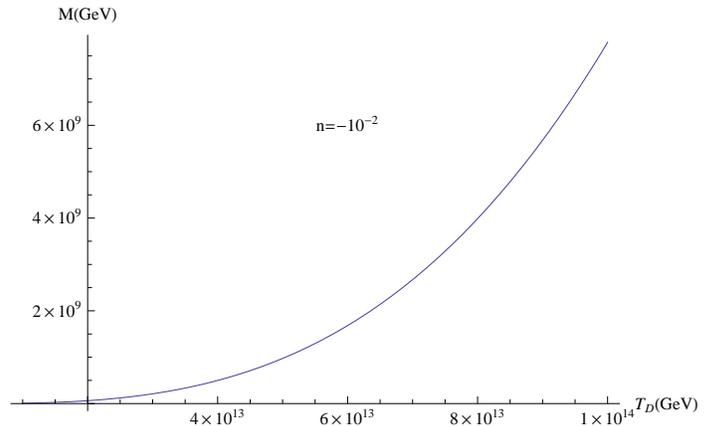}
\caption{\textit{HRN mass in $\ln R$ model in function of $T_D$.}} \label{eps67}
\end{figure}

\begin{figure}
\centering
\includegraphics[scale=1]{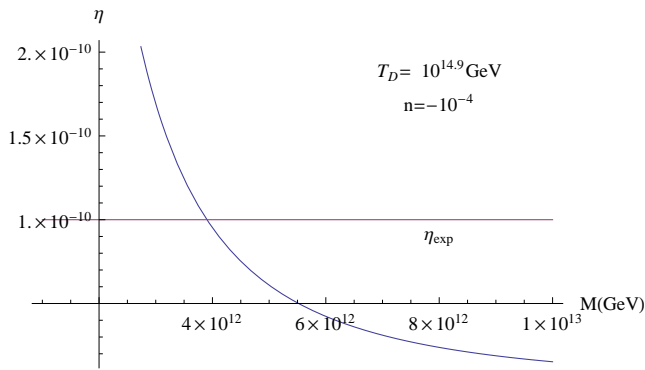}
\caption{\textit{LAF in function of HRN mass for $\ln R$ model  (\ref{etaleptoodin}).}}\label{eps5}
\end{figure}

Hereafter, we show the leptogenesis in the frame of the $ln (R)$ model of gravity, originally proposed here in this context. We proceed as it has been done before. We compute the same LAF given by (\ref{lamb2}) evaluating $\dot R$ from (\ref{frodin}), i.e., \begin{equation}
\eta=-\frac{ n (5 + 19 n + 22 n^2 + 8 n^3) \theta T_{}^{6}}{(2 + n)^2M^{2}M_{P}^3} \Bigg|_{T_{D}}.\label{etaleptoodin}
\end{equation}
We, at the beginning, impose   $n$ equal to  $-0.01$   and  $T_{D}=10^{16} GeV$. In this way LAF becomes:
\begin{equation}
\eta= \frac{5.9 \cdot 10^{23} GeV^2}{M^2}.
\end{equation}
In   Table
\ref{neutrino} we observe the variation of the mass of HRN
in function of the decoupling temperature $T_D$. In Figure
\ref{eps67} we show  $M$ {\it vs}  $T_D$.\\ Besides we exhibit results in agreement with the previous $f(R)$   model for $n=-10^{-4}$, observing that for $T_{D}=10^{15} GeV $ we get for the HRN a mass of the order of $10^{12} GeV$ in agreement with the see-saw mechanism prediction (\ref{seesaw}) as discussed before (LAF behaviour is in Figure \ref{eps5}).  Straightforwardly both $f(R)$ models introduced ($f(R)=R+\alpha R^2$ and $ln(R)$)  in order to explain leptogenesis give the expected LAF for the same value of the Temperature and HRN mass, i.e., $T_D\approx 10^{15} GeV$ and $M \approx 10^{12} GeV$ (how it is possible to see comparing results exposed in Figure \ref{eps3} and \ref{eps5}). This is another important hint that a little modification of the GR action is able to give the correct explanation of phenomenon such as leptogenensis.
\begin{table}
\centering
\begin{tabular}{|c|c|}\hline
$M$ & $T_{D}$ \\\hline
$10^{7.8}$ & $10^{13}$ \\\hline
$ 10^{10}$& $10^{14}$ \\\hline
$10^{14}$ & $10^{15}$ \\\hline
$10^{15}$ & $10^{16}$ \\\hline
\end{tabular} \caption{\textit{HRN mass and decoupling temperature in $GeV$ for $n=-0.01$ in $ln(R)$ gravities.}}\label{neutrino}
\end{table}

\section{Reconstructing  the  potential of a primordial scalar field from baryo-leptogenesis}
The extra gravitational degrees of freedom  generated from a $f(R)$ theory of gravity can be represented  from additional scalar fields. 
Here we  aim to find the correct potential form of a primordial scalar field that could be the main actor of leptogenesis and other phenomenon like Dark Energy or Inflation.

For this scope, let us  discuss the conformal transformation applied to a generic $f(R)$. Given a generic $f(R)$ action  it is possible, by  means of conformal transformation, to rewrite it in the Einstein frame, i.e., Ricci scalar plus a minimally coupled scalar field \cite{defelix, maeda, Gomez, Brans05}.

Hereafeter we consider the conformal transformation $e^{2 \chi}$ acting  on the metric
$g_{\mu\nu}$ \cite{defelix}, i.e.
 \begin{equation}\label{conftransf}
    {\tilde g}_{\mu\nu} = e^{2\chi}g_{\mu\nu}\,.
 \end{equation}
A particular choice of $\chi$ allow us to transform the beginning action $f(R)$ in a new one 
composed by Ricci scalar plus minimally coupled scalar field  \cite{dabrwconf, Faraoni2, Reconceiled}. 
In particular the right choice of $\chi$ is:
 \begin{equation}\label{chidef}
    \chi=\frac{1}{2}\ln |f'(R)|\,.
 \end{equation}
Imposing 
\begin{equation}\label{varphi}
    k \varphi = \chi\,, \qquad k=\frac{1}{\sqrt{6}}\,,
\end{equation}
the Lagrangian density of $f(R)$   can be rewritten in the (conformally) equivalent form \cite{defelix, pamela}
\begin{equation}\label{euivform}
    \sqrt{-g}f(R)=\sqrt{-{\tilde g}}\left(-\frac{1}{2}{\tilde R}+\frac{1}{2}\nabla_\mu \varphi \nabla^\mu \varphi - V\right)\,,
\end{equation}
where the potential $V$ is defined as
\begin{equation}\label{pot}
    V= \frac{f-R f'}{2f^{'\, 2}}\,.
\end{equation}

\begin{figure}
\resizebox{7.5cm}{!}{\includegraphics{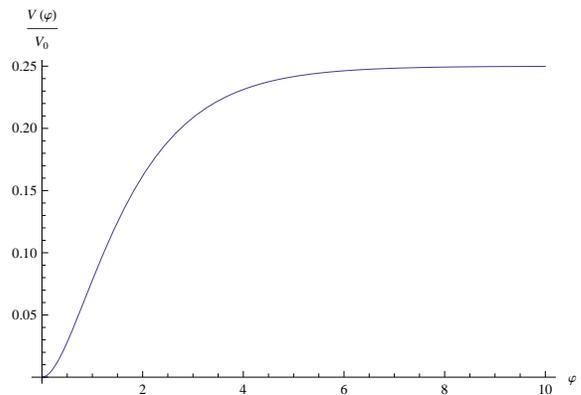}}
\caption{$\frac{V}{V_0}$ vs $\varphi$ for $f(R)=R+\alpha R^2$.}
 \label{fig1}
\end{figure}

\begin{figure}
\resizebox{7.5cm}{!}{\includegraphics{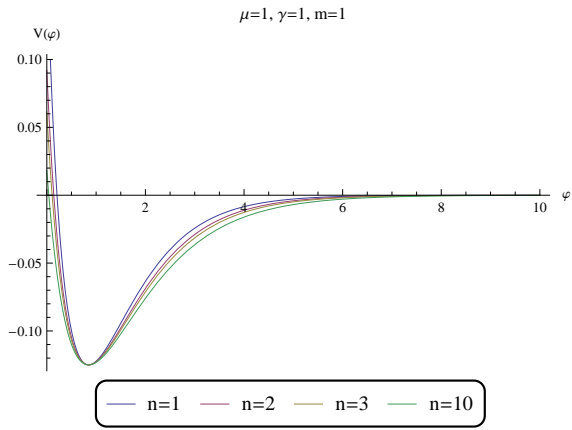}}
\caption{$V$ vs $\varphi$ for $f(R)=R+\gamma R^{-n}(ln \frac{R}{\mu^2})^m$ where $m, \gamma, \mu =1$. }
 \label{figall}
\end{figure}

\begin{figure}
\resizebox{7.5cm}{!}{\includegraphics{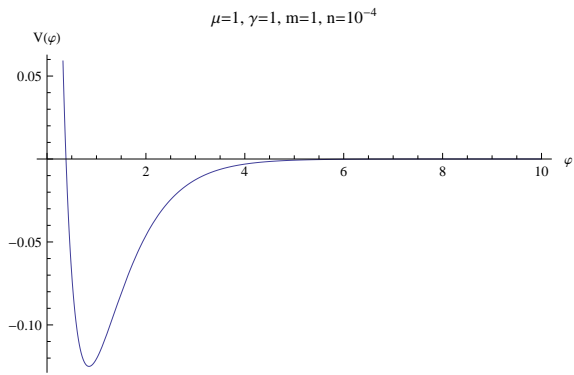}}
\caption{$V$ vs $\varphi$ for $f(R)=R+\gamma R^{-n}(ln \frac{R}{\mu^2})^m$ where  $m, \gamma, \mu =1, n=10^{-4}$. }
 \label{figall1}
\end{figure}

Thus, we can explicitly compute  the form of the potential $V$ in the case of $f=R+\alpha R^n$. From (\ref{chidef}), one gets
\begin{equation}\label{chiinvet}
    f' = e^{2k \varphi}\,,
\end{equation}
and  knowing that $f'=1+\alpha n R^{n-1}$, we can compute the Ricci in function of the new scalar field $\varphi$:
 \begin{equation}\label{Rvsphi}
    R=\left[\frac{1}{\alpha n}(e^{2k\varphi}-1)\right]^{\frac{1}{n-1}}\,.
 \end{equation}
Substituting this expression in (\ref{pot}), the potential  becomes
\begin{eqnarray} \label{pott}
  V  &=& \frac{2^{\frac{1}{n-1}}\alpha (1-n)}{(\alpha n)^{\frac{n}{n-1}}} e^{k \frac{4-3n}{n-1} \varphi}\left[\sinh k\varphi\right]^{\frac{n}{n-1}}\,.
\end{eqnarray}
In the case of $k\varphi\ll 1$ it is possible, by  means of Taylor expansion, to  assume a power law behaviour for the potential \cite{pamela}, i.e.
\begin{equation}\label{potV2}
    V\simeq V_0 \varphi ^\delta\,,
\end{equation}
with 
 \[
 V_0 \equiv \frac{2^{\frac{1}{n-1}}\,\alpha (1-n)}{(\alpha n)^{\frac{n}{n-1}}}\,,
 \quad \delta \equiv \frac{n}{n-1}\,.
 \]
 In our case, i.e., $n=2$ the potential (\ref{potV2}) assumes the following form:
 \begin{equation}
\frac{V}{V_0}=\varphi^2,
 \end{equation}
 where $V_0= - \frac{1}{2 \alpha}$.
While, always for $n=2$ the general expression for the potential (\ref{pott}) is
\begin{equation}
\frac{V}{V_0}=  e^{-2 k \varphi } \sinh ^2(k \varphi ),
\end{equation}
where $V_0= - \frac{1}{4 \sqrt{2} \alpha}$.  
The behaviour of such potential is in Figure \ref{fig1}.
\\
A more interesting potential form is obtained for the  $f(R)=R+\gamma R^{-n}(ln \frac{R}{\mu^2})^m$.
In the case of $m=1$ the Ricci in function of the scalar field is given by:
\begin{equation}
R=\left(\frac{\gamma  n W\left(\frac{\mu ^2 e^{1/n} (n+1) \left(e^{2 k \varphi }-1\right) \left(\mu ^2 e^{1/n}\right)^n}{\gamma  n}\right)}{(n+1) \left(e^{2 k \varphi }-1\right)}\right)^{\frac{1}{n+1}},
\end{equation}
where $W(z)$ denotes the function that gives the principal solution for $W$ in $z=We^W$.
The analytical expression of the potential $V(\varphi)$ becomes too complex (see Appendix A) and we report  its behaviour for different values of the parameter $n$ in Figure \ref{figall}. We fixed the other parameters, i.e., $\mu, \gamma$, because only $n$ is responsible for the determination of the scale factor of the Cosmos as  we have shown in Eq.   (\ref{aodin}). In  Figure \ref{figall1}, we represent the potential behaviour, for an order of magnitude of $n$ equal to the one previously used, in order to explain the baryo-leptogenesis, i.e. $n\approx 10^{-4}$. Proceeding in this way, we can claim that the $f(R)$  presents in  Eq. (\ref{frodin}), for $m=1$ and $n \approx 10^{-4}$, is  conformally equivalent to a scalar-tensor theory of gravity where the scalar field, minimally coupled to the Ricci scalar, has a potential  described  by  Figure \ref{figall1}.
Finally, it is worthwhile to point out that the general form of the two $f(R)$, i.e., (\ref{frmia}) and (\ref{frodin}), or their conformally equivalent  scalar-tensor theory with the potential described in this section, are able to generate baryo-leptogenesis, and they could also get rid of the problem of late and early time acceleration of the Cosmos.

\section{Analysis of the results and conclusions}

In this paper we have analysed  the long-standing problem of the origin of the asymmetry between matter and antimatter in the context of $f(R)$ theories. According to Occam's razor the solution here proposed, is the solution that needs fewer new conditions to solve the problems. In the context of $f(R)$ theories of gravity, we realize the correct  BAF/LAF  only in presence of 3 conditions: "CP violation, CPT dynamical violation, B/L violating reaction".  In particular the lepton asymmetry is generated by an interaction between chiral fermion and gravity (\ref{CPv}). This interaction splits energetic levels of neutrinos and antineutrinos  in an expanding Universe under the force of a $f(R)$ theory. Furthermore sphaleron converts this lepton asymmetry   to the baryon sector.
Besides,  we have also described baryogenesis in an alternative way,  in the context of an interaction  between baryon and Ricci scalar \cite{Stein} which reproduces the expected BAF as consequence of a little modification of gravity
\cite{LambM2}. 
 In this work we have proposed two different $f(R) $, i.e.
\begin{equation}\label{concl2}
f(R)=R+\alpha R^2,
\end{equation}
\begin{equation}\label{concl4}
f(R)=R+\gamma
R^{-n}\left( \ln \frac{R}{\mu^2}\right)^m,
\end{equation}
where, in particular, the second one has never been studied before in the context of the matter-antimatter asymmetry problem.

Both model satisfy the conditions enumerated hereafter, i.e.
\begin{enumerate}
\item $\epsilon \ll 1$ in order to not change the standard thermal history of the Universe.
 \item reproduce LAF and BAF consistent with experimental data.
 \end{enumerate}
 
Besides we have found the potential  of a primordial scalar field, trough LAF constraints. Actually, $f(R)$  gravities are conformally equivalent to a theory with traditional Einstein term plus a scalar field. It is possible to find the  potential, for the scalar-tensor theory equivalent to the $f(R)$, trough conformal transformation of the metric. For both $f(R)$ analysed in this paper we have found the potential of the scalar field generating the asymmetry between matter and antimatter. 
 
 It is worthwhile to highlight how $f(R)$ theories of gravity can introduce a small perturbation to the GR scale factor that allows us to obtain the expected value for the asymmetry factor.    
   
   At the end we point out that, as shown in this paper, $f(R)$ theories may be the ultimate solution for most of open problems in modern cosmology, e.g. Dark Energy, Inflation, Dark Matter, Bario-Leptogenesis.

\section*{Acknowledgements}
We wish to thank I.N.F.N for supporting our studies, A. Strumia, G. Lambiase, V.Galluzzi for useful discussions  and E. Vicari.

\begin{widetext}

\appendix
\newpage
\textbf{Appendix A}

The general expression for the potential of the scalar field action conformally equivalent to $f(R)=R+\gamma R^{-n} ln(\frac{R}{\mu^2})$ is:
\begin{equation}
V(\varphi)= \frac{A}{B},
\end{equation}
where 
\begin{equation}
B=2 \left(\gamma +\left(\left(\frac{\gamma  n W\left(\frac{(n+1) \left(e^{2 k \varphi }-1\right) \left(\mu ^2 e^{1/n}\right)^{n+1}}{\gamma  n}\right)}{(n+1) \left(e^{2 k \varphi }-1\right)}\right)^{\frac{1}{n+1}}\right)^{n+1}-\gamma  n \log \left(\frac{\left(\frac{\gamma  n W\left(\frac{(n+1) \left(e^{2 k \varphi }-1\right) \left(\mu ^2 e^{1/n}\right)^{n+1}}{\gamma  n}\right)}{(n+1) \left(e^{2 k \varphi }-1\right)}\right)^{\frac{1}{n+1}}}{\mu ^2}\right)\right)^2
\end{equation}

and
\begin{eqnarray}
A =& \left(\gamma  \left((n+1) \log \left(\frac{\left(\frac{\gamma  n W\left(\frac{(n+1) \left(e^{2 k \varphi }-1\right) \left(\mu ^2 e^{1/n}\right)^{n+1}}{\gamma  n}\right)}{(n+1) \left(e^{2 k \varphi }-1\right)}\right)^{\frac{1}{n+1}}}{\mu ^2}\right)-1\right)\right. \nonumber \\ &\cdot \left.\left(\frac{\gamma  n W\left(\frac{(n+1) \left(e^{2 k \varphi }-1\right) \left(\mu ^2 e^{1/n}\right)^{n+1}}{\gamma  n}\right)}{(n+1) \left(e^{2 k \varphi }-1\right)}\right)^{\frac{2}{n+1}} \left(\left(\frac{\gamma  n W\left(\frac{(n+1) \left(e^{2 k \varphi }-1\right) \left(\mu ^2 e^{1/n}\right)^{n+1}}{\gamma  n}\right)}{(n+1) \left(e^{2 k \varphi }-1\right)}\right)^{\frac{1}{n+1}}\right)^n\right).
\end{eqnarray}
$W(z)$ is the function that gives the principal solution for $W$ in $z=We^W$ and in \emph{Wolfram Mathematica\textsuperscript\textregistered} is named $ProductLog[z]$.
\end{widetext}
\end{document}